\documentclass[aps,prb,twocolumn,amsmath,amssymb,superscriptaddress,citeautoscript,10pt]{revtex4-2}
\usepackage{graphicx}
\usepackage{dcolumn}
\usepackage{bm}
\usepackage{latexsym,epsfig}
\usepackage{graphicx}
\usepackage[
    format=plain,
    figurename=Figure
]{caption}

\usepackage{ragged2e}
\DeclareCaptionJustification{plain}{\justifying}
\usepackage[subrefformat=simple,labelformat=simple]{subcaption}

\usepackage[export]{adjustbox}
\usepackage{verbatim}
\usepackage{comment}
\usepackage{amsmath}
\usepackage{amssymb}
\usepackage{mathrsfs}
\usepackage{stmaryrd}
\usepackage{xcolor}
\usepackage{epstopdf}
\usepackage{grffile}
\usepackage{mathtools}
\usepackage{physics}
\usepackage{ulem}
\usepackage{siunitx}
\AtBeginDocument{\RenewCommandCopy\qty\SI}
\DeclareSIUnit\hopping{\mathit t_0}

\DeclareTextFontCommand{\emph}{\itshape}
\DeclareGraphicsExtensions{.eps}

\usepackage{hyperref}
\hypersetup{
    colorlinks=true,      
    urlcolor=blue,
    citecolor=blue,
    linkcolor=blue
}

\usepackage[capitalize]{cleveref}
\crefname{figure}{Fig.}{Figures}

\begin{document}
\title{Breakdown of bosonic Thouless pump  due to interaction in a quasiperiodic lattice} 
\author{Suman Mondal}
\affiliation{Max Planck Institute for the Physics of Complex Systems, N\"{o}thnitzer Str. 38, 01187 Dresden, Germany}
\author{Emmanuel Gottlob}
\affiliation{Cavendish Laboratory, University of Cambridge, JJ Thomson Avenue, Cambridge CB3 0HE, United Kingdom}
\author{Fabian Heidrich-Meisner}
\affiliation{Institut f\"{u}r Theoretische Physik, Georg-August-Universit\"{a}t G\"{o}ttingen, D-37077 G\"{o}ttingen, Germany}
\author{Ulrich Schneider}
\affiliation{Cavendish Laboratory, University of Cambridge, JJ Thomson Avenue, Cambridge CB3 0HE, United Kingdom}

\date{\today}

\begin{abstract}
We investigate the effect of inter-particle interaction on the quantized Thouless pump in the bosonic quasiperiodic Aubry-Andr{\'e} model and find that the quantization of the pumped charge breaks down already for weak interactions.  Furthermore, the pumped charge undergoes sharp changes as a function of interaction strength that we can attribute to the closing of specific doublon channels. As expected, the quantization revives in the hard-core limit at very large interaction strengths where the bosons are subject to a hardcore constraint. 
Interestingly, the stability of isolated doublons under the pump depends on the band they are in. For repulsive interactions and a suitably fixed pump period, doublons in the lowest band are pumped stably while doublons in higher bands dissociate during the pump with one particle decaying into a lower band. This asymmetry leads to the decay of the total energy over time, in stark contrast to the typical Floquet heating expected for a driven many-body system.

\end{abstract}

\maketitle
\section{Introduction}

The study of topological properties of matter reveals fundamental phenomena such as robust quantized transport~\cite{Thouless1982,Halperin1982,Hasan2010}. Topological Thouless pumps, time-periodic analogs of the two-dimensional Chern insulators, have emerged as a cornerstone for exploring quantized adiabatic transport in low-dimensional systems~\cite{Thouless1983, MonikaRev2023}. While the topological nature of noninteracting periodic systems is well classified~\cite{Hasan2010}, a significant effort has been directed towards 
establishing a conceptual understanding of topology and transport in periodically driven 
quasiperiodic systems in the last two decades~\cite{Verbin2013,Kraus2013,Goldman2015,Bandres2016,He2019}. In the context of the Thouless pump, the Aubry-Andr{\'e} model exhibits quantized pumping when the phase of the potential is adiabatically changed with time~\cite{Kraus2012,Emmanuel2024,Pasquale2020}. This topological response originates from the topology of the underlying energy bands.

\begin{figure}
\centering
    \includegraphics[width=1\linewidth]{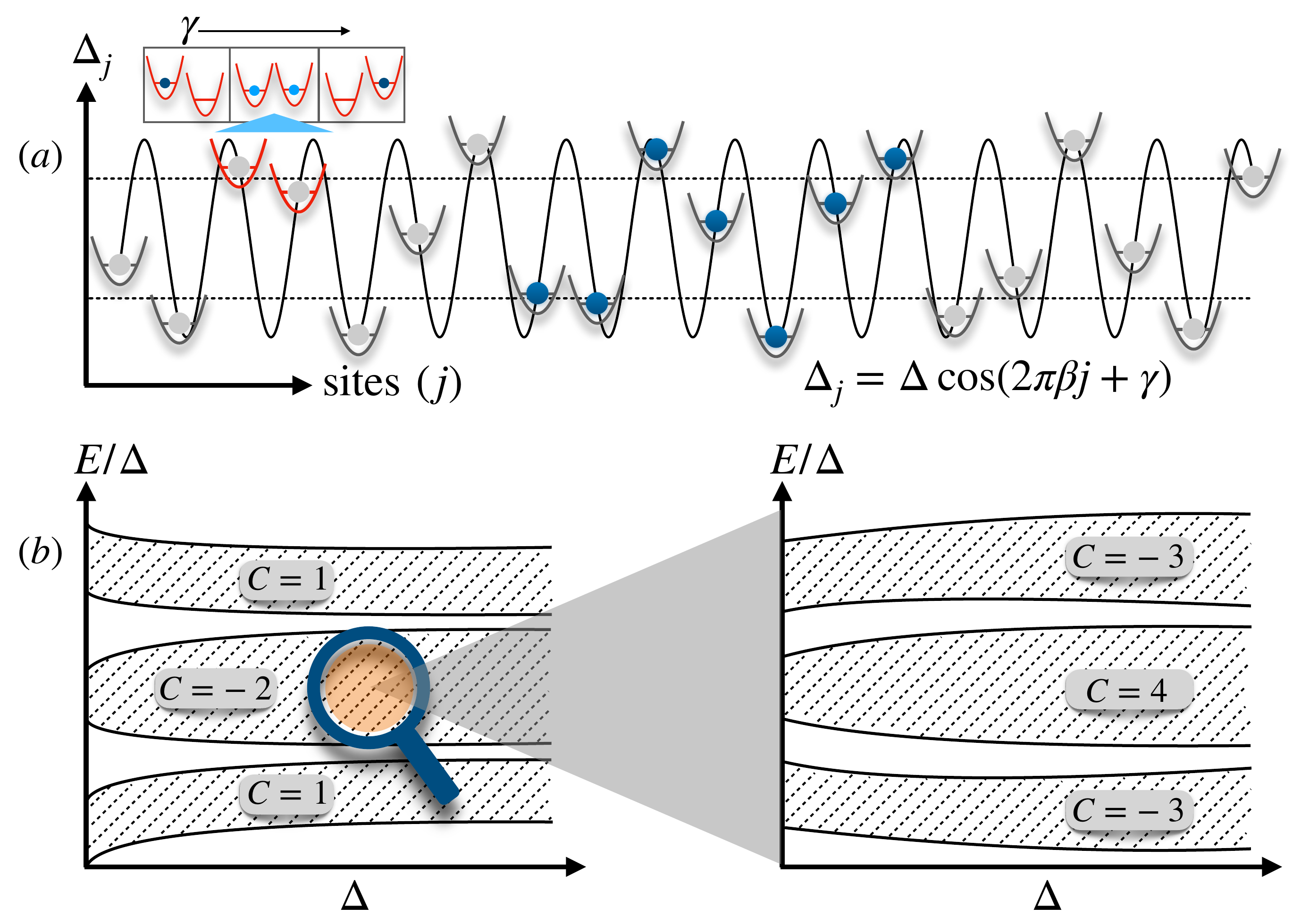}
\caption{(a) Potential energy landscape (in arbitrary units with $\beta = \sqrt{2}/2$) of a one-dimensional Aubry-Andr\'e model defined in Eq.~\eqref{eq:hamiltonian}. The parabolae show the potential minima of the lattice potential at different sites ($j$). The circles represent the sites and the  quasiperiodic onsite energy ($\Delta_j$), and the blue shaded circles depict singly-occupied sites in the initial state. The two sites highlighted in red illustrate a resonance. As depicted in the inset, the variation of  $\gamma$ during the pump gives rise to a Landau-Zener transition that pumps the particle from left to right and vice versa. The dotted line marks the position of the center of the first-order gaps in the single-particle spectrum. (b) Illustration of the hierarchical structure of gaps in the single-particle spectrum. The first-order gaps define the three bands (left) with different Chern numbers ($C$) and the 
higher-order gaps successively divide the bands into a fractal of  sub-bands (second-order gaps shown on right).}
\label{fig:intro}
\end{figure}
Recently, there has been a resurgence of interest in quasiperiodic systems to study the Bose glass~\cite{Chiara2014,Ciardi2023,Emmanuel2023,Yu2024}, Anderson localization and mobility edge~\cite{DasSarma2017,Ulrich2018,Wang2020,Ulrich2020,DasSarma2020,Mishra2021,Mishra2022}, many-body localization~\cite{Shankar2013,Sarang2017,Ulrich2017,Strkalj2022,Steven2023}, and Thouless pumping~\cite{Takahashi2021pump, Emmanuel2024, Moustaj2025}, to name some key examples. 

Interactions can alter the properties of the Thouless pump. In some instances, interactions can give rise to new pumping regimes or even induce charge pumping that would not occur otherwise~\cite{Eric2024,Hatsugai2020,sumantopo1,mondal2021,Bertok2022,mondal2022,suman2024,Rajashri2024,Jurgensen2025,Chaudhari2025}, while in many other scenarios, they can disrupt the quantized pump~\cite{ArguelloLuengo2024,Konrad2023,Nakagawa2018,Hayward2018,Suman2022}. To date, research in this area has primarily focused on periodic systems. On the one hand, the effect of interactions on Thouless pumping in quasiperiodic systems remains largely unexplored. On the other hand, advances in ultracold atomic gas setups have enabled experimental realizations of topological pumps~\cite{Lohse2016,Nakajima2016,Schweizer2016,Lohse2018,Takahashi2021pump,Konrad2023,Eric2024,Konrad2024}. Notably, topological pumping in the presence of interactions has recently been achieved in experiments, where interactions play a pivotal role in the charge transport~\cite{Konrad2023,Eric2024,Konrad2024}. Furthermore, quasiperiodic lattices in these experimental setups are also feasible~\cite{Roati2008,BlochMBL2015,Takahashi2021pump}. Motivated by these facts, we study the effect of interactions on Thouless pumping in quasiperiodic systems.

The starting point for our work is a recent investigation of Thouless pumping of noninteracting particles in the quasiperiodic  Aubry-Andr{\'e} chain by  Gottlob \textit{et al.}~\cite{Emmanuel2024} 
(including some of the present authors) elucidating the intricate relationship between the pumping speed and the hierarchical order of gaps in the spectrum.
For strong quasiperiodic potentials, the eigenstates are strongly localized, and a hierarchical order of gaps arises from higher-order tunneling processes connecting distant resonant sites. During the pump protocol, which consists of sliding the quasiperiodic modulation across the lattice, Landau-Zener transitions between these resonant sites facilitate the pumping of particles.
 Due to this hierarchy of increasingly smaller gaps, a truly adiabatic pump is impossible in such quasiperidodic systems. Rather, the pumping period $T$ can be chosen such that  its inverse is smaller than
all gaps up to a certain order, but much larger than all smaller ones. 
A key finding was the surprising resilience of quantization in the Thouless pump, notably persisting beyond the disorder-induced closure of spectral gaps. This gap closure conventionally signifies a breakdown condition in periodic systems~\cite{Takahashi2021pump,Hayward2021,Cerjan2020,Wauters2019,Wang2019,Imura2018}.

To analyze the effect of interactions, we consider a pump period where the pump is approximately adiabatic up to the first-order energy gaps that are the largest gaps in the non-interacting system. As the initial state we employ a product state where the central sites are singly occupied by bosons, see Fig.~\ref{fig:intro}, and time evolve it according to the pump protocol of~\cite{Emmanuel2024}, see Eq.~\eqref{eq:pump}. We find that the interaction between particles significantly affects the pump. Most importantly, even weak interactions can cause a  breakdown of quantized pumping. Conversely, in the case of very strong interaction, the quantized Thouless pump is revived as expected due to the emergence of the hardcore limit, where the system can be mapped to the non-interacting fermions. In the  regime of intermediate interaction strengths, we observe some sharp changes in the pumped charge that we can trace back to the closing of specific doublon channels. 

Furthermore, we  analyze the pumping of isolated doublons and uncover an interesting difference in the stability of the doublon pump depending on the initial energy. 
For repulsive interactions, doublons in the lowest band are pumped stably while doublons in higher bands dissociate, transmitting one particle into a lower band. This phenomenon results in an unusual energy decay in the driven many-body system, where the system continuously loses energy. This is in stark contrast to the usual Floquet heating that is accompanied by increase of energy in a periodically driven many-body system.

In the remainder of the paper we discuss the theoretical model considered in this study in Sec.~\ref{sec:model}, along with the method used for the time evolution. In Sec.~\ref{sec:manybody}, we present the effect of interactions on the pumped charge in the many-body setting and discuss the role of individual doublon channels. We explore the asymmetry in the stability of a doublon pump depending on the energy of an isolated doublon in Sec.~\ref{sec:2particle} and its effect on total energy in Sec.~\ref{sec:enrdk}. Finally, we conclude in Sec.~\ref{sec:conclusions}.

\begin{figure*}
\centering
    \includegraphics[width=1\linewidth]{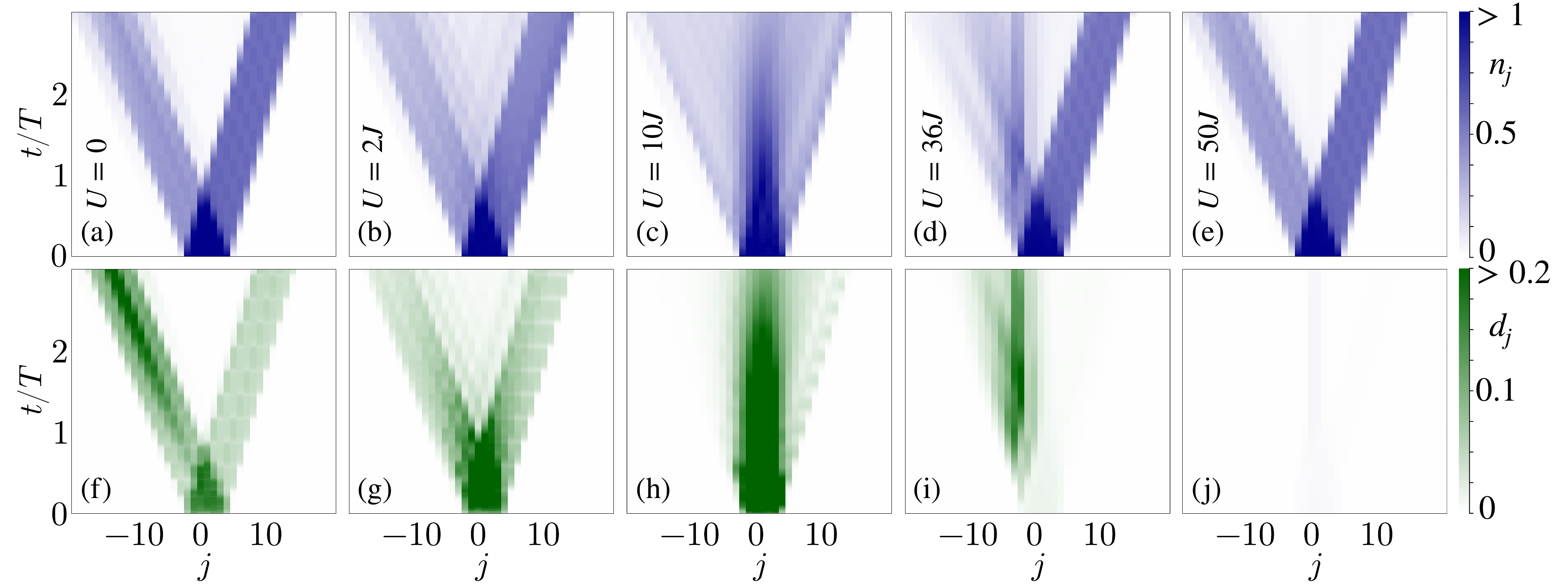}
\caption{Time evolution of the particle ($n_i$) and doublon density ($d_i$)  during the pump for $N=7$, $\Delta = 15J$, and for different values of $U$ shown in the upper and lower panels, respectively.  For $U=0$, the initial cloud bifurcates into two branches with time, corresponding to the $C=-2$ and $C=1$ bands. For $U=2J$, $10J$, and $36J$, the time evolution of the density
changes drastically compared to $U=0$. At very large $U=50J$, the time evolution of the density  appears to be similar to the case of $U=0$. The time evolution of the doublon density  is distinctly different in all cases (see the main text). System size is $L=40$ and results are averaged over $40$ realizations (i.e., 40 different values for $\gamma(0)$).}
\label{fig:7particles}
\end{figure*}

\section{Model and method}\label{sec:model}
The model Hamiltonian that describes our quasiperiodic interacting system consists of three terms,
\begin{align}
    \label{eq:hamiltonian}
    \hat H &= \hat H_\text{kin} + \hat H_\text{qp}+ \hat H_\text{int},
\end{align}
where $\hat H_\text{kin}$ is the kinetic energy term, $\hat H_\text{qp}$ is the quasiperiodic potential term and $\hat H_\text{int}$ represents the interaction between the bosonic particles: 
\begin{eqnarray}
 \hat H_{\text{kin}} &=& -J \sum_{l=1}^L (\hat b^\dagger_{l} \hat b_{l+1} + \mathrm{h.c.})\,,\\
 \hat H_{\text{qp}} &=& \Delta \sum_{l=1}^L \cos({2\pi\beta l + \gamma(t)}) \hat n_{l}\,,\\
 \hat H_\text{int} &=&  \frac{U}{2} \sum_{l=1}^L \hat n_{l}(\hat n_{l} -1) \,.
\end{eqnarray}
Here, $\hat b_{l}$ and $\hat n_{l} = \hat b_{l}^\dagger  \hat b_l$ are the bosonic annihilation and density operators on site $l$ in a chain with $L$ sites. $J$ is the nearest-neighbour tunneling matrix element, $\Delta$ represents the strength of the quasiperiodic potential with irrational period $\beta = \sqrt{2}/2$, and $U$ denotes the on-site two-body interaction energy. We use $J=1$, which defines the unit of energy. 

In this setup, pumping consists of linearly increasing the phase of the quasiperiodic modulation $\gamma(t)$ with period $T$ such that $\gamma(t+T)=\gamma(t)$, i.e.:
\begin{equation}\label{eq:pump}
    \gamma(t)=\gamma(0)+t\,2\pi/T \,.
\end{equation}
Previous work~\cite{Emmanuel2024} by some of us contains a detailed discussion of the non-interacting pump in the  strongly-localized regime at large $\Delta$. As illustrated in Fig.~\ref{fig:intro}(b), in this case, the single-particle spectrum contains a hierarchy of exponentially decreasing energy gaps, $\delta E_{(k)} \propto J^k/\Delta^{k-1}$. These gaps correspond to resonances between two sites that are $k$ sites apart. Due to the exponentially decreasing size of the gaps, one can set a suitable pump period $T$ so to be adiabatic for all gaps up to a particular order, while being fast compared to all smaller gaps.  
The first-order gaps $k=1$ separate three energy bands with Chern numbers $C = 1$, $-2$, and $1$ as illustrated in Fig.~\ref{fig:intro}(b). Smaller gaps further divide these primary bands into sub-bands with higher Chern numbers in a hierarchical structure [see Fig.~\ref{fig:intro}(b)]. During pumping, an isolated atom will remain at its site until $\gamma(t)$ is such that its site becomes resonant with its $k$-th neighbor, see red sites in Fig.~\ref{fig:intro}(a) for $k=1$, at which point it will transition to this resonant neighboring site following an approximately adiabatic Landau-Zener transition.

We confine our analysis to a fixed period $T= \frac{2 \pi}{0.05J}$ and large $\Delta = 15J$, where the first-order gaps dominate pumping~\cite{Emmanuel2024} and study the effect of interactions on the pump. 

As our initial state, we consider a ($n=1$) product state that contains particles only in the center of the lattice. The initial state for $N$ particles can be written as,
\begin{equation}\label{eq:psi0}
    |\psi_N(0)\rangle = \left( \prod_{l=\frac{L}{2}-\frac{N}{2}}^{\frac{L}{2}+\frac{N}{2}} \hat{a}^\dagger_l \right) |{\rm{vac}}\rangle.
\end{equation}
A pictorial representation is shown in Fig.~\ref{fig:intro}(a). Because of the  large $\Delta = 15J$, the particles are highly localized  and hence there is no significant expansion of the cloud without the drive. Hence, the particles can essentially only move because of the periodic drive during the pump. Note that the initial state overlaps with states in all bands of the single-particle spectrum.

We use the matrix-product-state (MPS) based time-evolving block decimation (TEBD) method to time evolve the initial product states~\cite{Vidal2004,Schollwock2011,Paeckel2019}. In the TEBD method, the time-evolution operator is Suzuki-Trotter decomposed in second order, producing an error $O(\delta t^3)$ per time step ($\delta t$) of evolution which we consider to be $\delta t = 0.01/J$. This is implemented using the ITensor library~\cite{itensor} and the maximum bond dimension is set to $1500$, which is never reached for the truncation error of $10^{-8}$ and for the times simulated in this paper. We truncate the local bosonic Hilbert space to $n\leq4$. To mitigate any dependence on specific values of the phase $\gamma$, the results are averaged over $40$ independent realizations, achieved by uniformly sampling the initial phase $\gamma(0)$ from the interval $[0, 2\pi]$.

\begin{figure}
\centering
\includegraphics[width=1\linewidth]{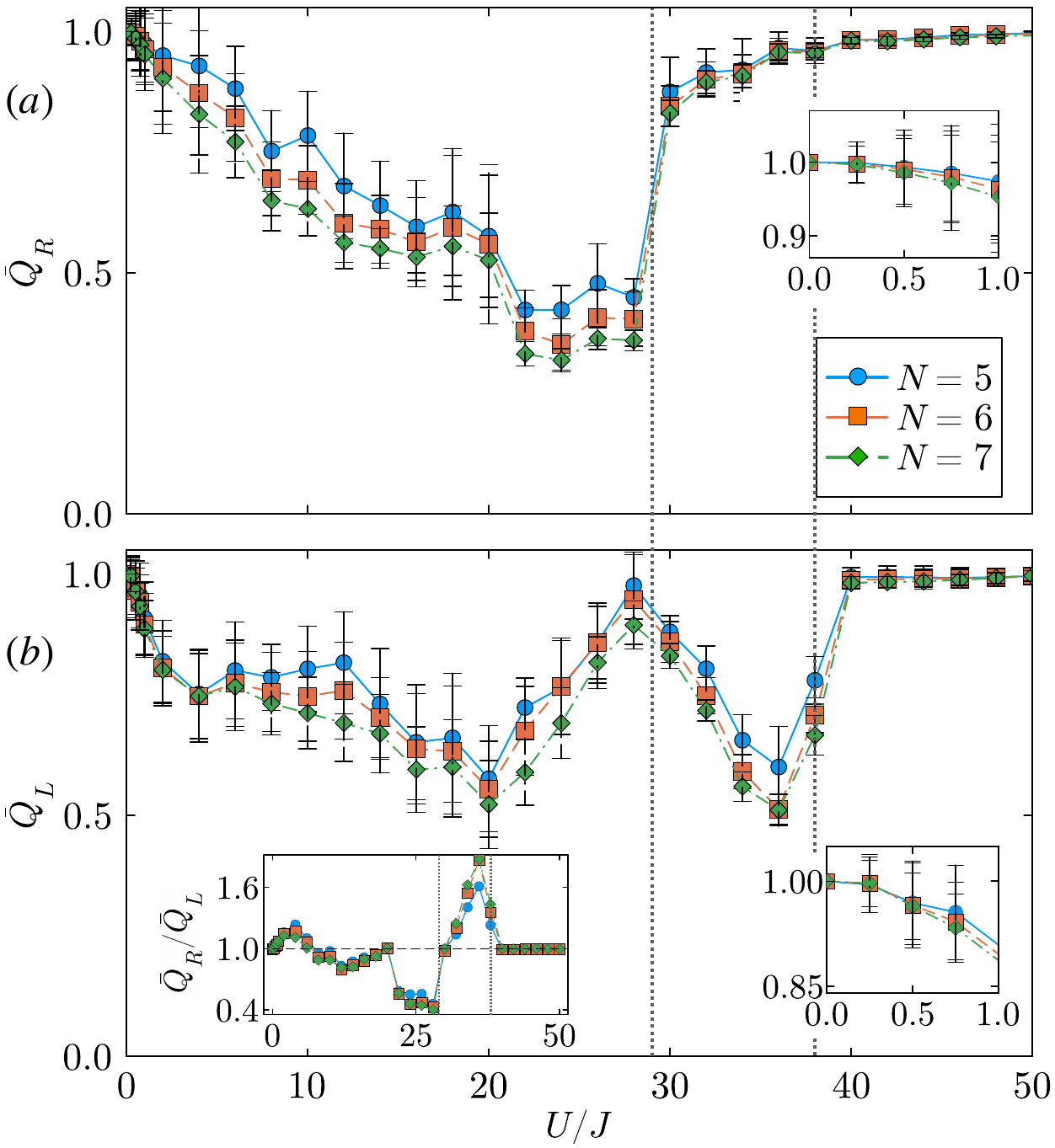}
    \caption{Pumped charge in the time window from $t_1=2T$ to $t_2 = 3T$, normalized by the non-interacting case,  as a function of interaction strength for $\Delta = 15J$ for (a) the right and (b) the left half of the system. The simulations employ a system size of $L=40$ sites, results are averaged over $40$ initial values of $\gamma(0)$, and are shown for three different particle numbers $N=5$, $N=6$, and $N=7$. The right insets show a magnified view of the weak-interaction regime, and the left inset in (b) shows how the ratio of $\bar{Q}_R/\bar{Q}_L$ changes as a function of  interaction strength. The vertical dotted lines denote the values of $U$ where the system undergoes a sharp change in its behavior due to the resonances discussed in the text.}
\label{fig:QLR}
\end{figure}

\begin{figure*}
\centering
\includegraphics[width=1\linewidth]{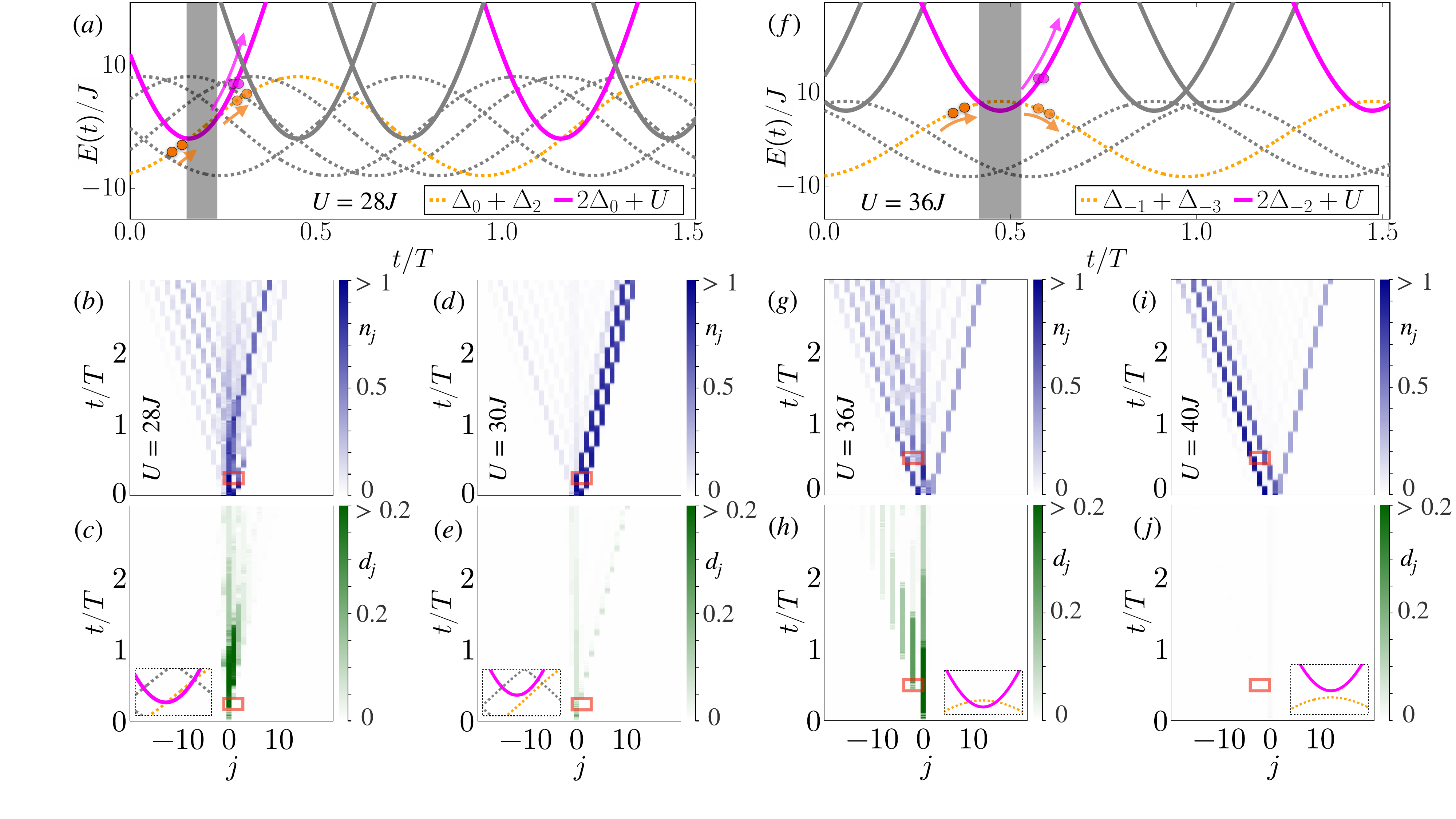}
\caption{The figure is divided into two parts, (a)-(e) and (f)-(j), corresponding to the dotted lines at $U_1\sim29J$ and $U_2\sim38J$ in Fig.~\ref{fig:QLR}. In both cases we examine values of $U$ on both sides of the lines shown in Fig.~\ref{fig:QLR} to highlight the contrast in the dynamics between a doublon channel being open or closed. (a) Solid lines denote the energy of a doublon on site $j$ and dotted lines correspond to pairs of singlons on sites $j$ and $j+2$. 
The initial phase of the quasiperiodic potential is $\gamma(0) = 0.4\pi$ and $U=28J<U_1$. (b) and (c) depict the evolution of particle ($n_j$) and doublon density ($d_j$) starting from a pair of particles on sites $j=0$ and $j=2$, corresponding to the orange dotted line in (a). The shaded area in (a) and red boxes in (b) and (c) indicate the time window where the energy of the initially occupied channel crosses with the doublon on site $j=0$.
(f)-(j) Analogous plots around $U\sim 38J=U_2$ and  $\gamma(0) = 0.6\pi$ with two particles initially on sites $j=-1$ and $j=1$. For $U=36J<U_2$, long-lived doublons form in the left half of the system, but this is strongly suppressed for $U=40J>U_2$.
}
\label{fig:resonance}
\end{figure*}

\section{Many-body pump}\label{sec:manybody}

\subsection{Dynamics of density and doublon density}

To probe the influence of interactions on the pumping, we time evolve the initial state of~Eq.~(\ref{eq:psi0}) with the pump protocol in~Eq.~(\ref{eq:pump}). Figure~\ref{fig:7particles} presents the time evolution of the local density $n_j=\langle \hat n_j \rangle$ and doublon density $d_j = \frac{1}{2} \langle \hat n_j(\hat n_j - 1) \rangle$. 
In the non-interacting limit [Fig.~\ref{fig:7particles}(a) and (f)], the initial  cloud  splits into two distinct branches that directly indicate quantized Thouless pumping of particles in bands with different topological character: the right-moving lower and upper bands with  $C=1$ and the left-moving middle band with $C=-2$~\cite{Emmanuel2024}. While the initial product state contains zero doublons, tunneling creates a finite number of doublons already within the first tunneling time~\cite{Ronzheimer2013}. Comparing the non-interacting case with the case of low interaction strength, $U=2J$ [Fig.~\ref{fig:7particles}(b) and (g)], we observe a difference in the time evolution of the density. We see a finite particle density in between the two branches, indicating back-scattering due to the interactions. For stronger interactions $U=10J$ [Fig.~\ref{fig:7particles}(c) and (h))], a high-density core remains longer and keeps spreading slowly over time.
At the even higher interaction strength $U=36J$ [Fig.~\ref{fig:7particles}(d) and (i)], the left-moving branch is strongly affected by the formation of  long-lived doublons discussed in more detail below, while the right-moving branch is largely unaffected by the interaction. Finally, for very large interaction strength such as $U=50J$ [Fig.~\ref{fig:7particles}(e) and (j)], the time evolution of the density becomes similar to the non-interacting case and the pumping becomes quantized again. This signals the onset of the limit of hardcore bosons, where dominant interactions fully suppress the formation of higher occupancies~\cite{Girardeau1960,Rigol2005,Ronzheimer2013}.

\subsection{Pumped charge}
For a more quantitative analysis, we turn to the pumped charge, that is, the number of particles being pumped through bond $j$ of the $1$D lattice in the time window between $t_1$ and $t_2$. This is done by integrating the current $I_j(t) = -2J \Im  \langle \Psi(t)| \hat{b}_{j+1}^\dagger \hat{b}_j |\Psi(t)\rangle$ on that bond using
\begin{equation}\label{eq:Qt}
    q_j(t_1,t_2) = \int_{t_1}^{t_2} I_j(t) \,dt \,.
\end{equation}
This calculation is performed over a full pumping period, specifically between times $t_1=2T$ and $t_2=3T$. 
 To isolate and clearly identify the contributions to the pumped charge from the $C=1$ and $C=-2$ bands, which are pumped in opposite directions for  $U=0$, the pumped charge $q_{j}$ is summed separately over the right ($Q_R$) and left half ($Q_L$) of the system.
Thus, we define
\begin{equation}\label{eq:Q12}
    Q_{R/L} = \sum_{j\in R/L} q_j \,,
\end{equation}
where $j$ belongs to the right ($R$) or left ($L$) halves of the system. We normalize  $Q_R$ and $Q_L$ at finite $U$ by their correspondind values in $U=0$ case via
\begin{equation}\label{eq:Q12b}
\bar{Q}_{R/L}(U) = \frac{Q_{R/L}(U)}{Q_{R/L}(U=0)}.  
\end{equation}
For a large number of particles in the initial state, $Q_R (U=0) = -Q_L(U=0)$.

The results are presented in Fig.~\ref{fig:QLR} and show that already small values of $U$ are sufficient to reduce the pumped charge and break its quantization, see also the right insets in Fig.~\ref{fig:QLR}(a) and (b).
In this regime, the interaction modifies the local level structure and the energy gaps in the two-site Landau-Zener picture with two particles, thereby modifying the adiabaticity condition and potentially leading to a non-adiabatic transition. The interaction effect hence quantitatively depends on the chosen pumping speed, and could, for a single isolated Landau-Zener problem, be cured by reducing the pumping speed. For the full quasiperiodic problem, however, this is not possible since higher-order resonances with $k>1$ become relevant for slower pumps~\cite{Emmanuel2024}. This indicates that the pump is generically not robust against on-site interactions, in stark contrast to its exceptional robustness against disorder potentials \cite{Emmanuel2024}. 

For intermediate interactions, the left- and right-moving bands are influenced differently by the interactions, leading to unequal currents with $\bar{Q}_R/\bar{Q}_L\neq1$. The left inset in Fig.~\ref{fig:QLR}(b) demonstrates that this ratio depends sharply on interactions  - see below.  

For very large interactions ($U>40J$), once $U$ dominates over both kinetic ($J$) and potential energy $\Delta$, the interaction effects diminish as the system enters the hardcore regime, where double occupancies are suppressed, and the pumped charges approach the quantized values of the non-interacting limit. This situation can be compared to the
stability of a Thouless pump in a bosonic system without disorder potentials:
coming from the hardcore-boson limit, the pump remains quantized until the superfluid phase is encountered
and the system becomes gapless
upon lowering the strength of a Hubbard interaction
\cite{Hayward2018}. The breakdown occurs in a different way compared to the situation studied in our work.

\begin{figure}
\centering
\includegraphics[width=1\linewidth]{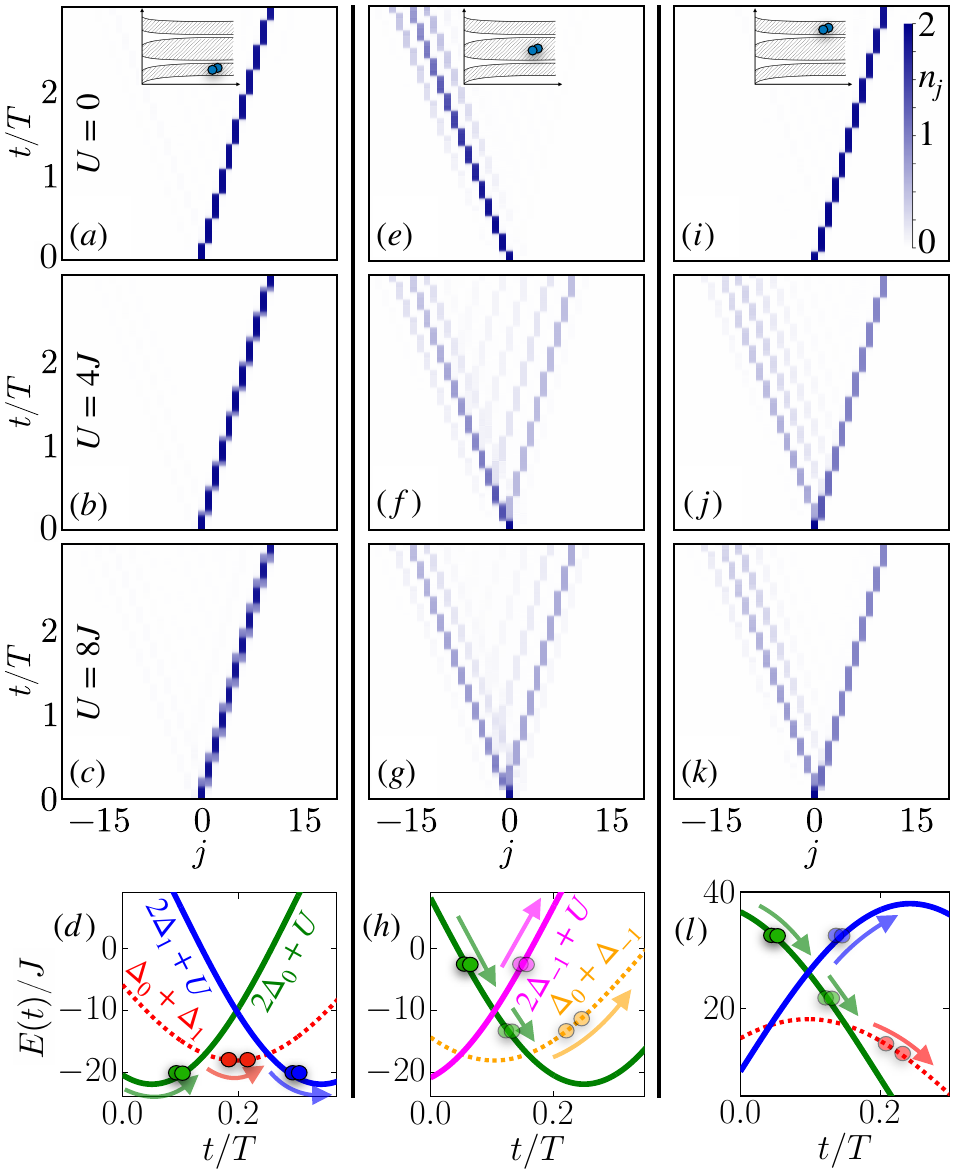}
\caption{Illustration of  the asymmetry in the  stability of the  doublon pump depending on its initial energy. The time evolution of the density is plotted for different values of $U$: $U=0$, $U=4J$, and $U=8J$. Here, two particles are initially placed at the central site for three initial values of $\gamma (0) = 1.2\pi$ (a-c), $0.8\pi$ (e-g), and $0.4\pi$ (i-k) that correspond to the energies belonging to lower ($C=1$) middle ($C=-2$), and upper ($C=1$) bands, respectively, for $U=0$. We see that, in the presence of interactions, the pumping of doublons  in the lowest band remains stable, whereas it is unstable for the middle and upper band. To explain the presence or absence of stability in each case, we plot the energy channels formed by local doublon (solid lines) and singlon (dotted lines) states in (d), (h), and (l) for the parameters considered in (c), (g) and (k) respectively. The arrows represent the path followed by the initial state during the evolution. In the first case (lower band), the doublon successfully adiabatically moves from one site to the next site via a singlon state. In the other two cases (middle and upper band), the dynamics becomes non-adiabatic and the doublon fails to move to the other site.
}
\label{fig:2particle}
\end{figure}

\subsection{Closing of doublon channels}
For intermediate interaction strength ($20J\lesssim U \lesssim 40J$), the left- and right-moving parts exhibit distinct behaviors, giving rise to unequal left- and right-flowing currents, see the left inset in Fig.~\ref{fig:QLR}(b). 
In particular, there are several sharp changes in the pumped charges, marked by the dotted lines at $U_1\approx29J$ and $U_2\approx38J$ in Fig.~\ref{fig:QLR}. As we will argue below, these abrupt changes are caused by the closing of specific doublon channels. 

In Fig.~\ref{fig:resonance}(a)-(e), we focus on the first sharp change at around $U_1\approx 29J $, at which $\bar{Q}_R$ markedly increases. 
We use a minimal setup with two particles starting on sites $j=0$ and $j=2$. Figure~\ref{fig:resonance}(a) depicts the total energy of the relevant channels: Dotted lines represent the energies of pairs of singlons at sites $j$ and $j+2$ and  solid lines represent the total energy of doublon states when both particles occupy the same site. Notably, for $U<U_1$, these lines cross (shaded area) during the pump cycle such that the pumping can convert a pair of singlons into a doublon. 
Figures~\ref{fig:resonance}(b) and (c) show the time evolution of the particle $n_j$ and doublon densities $d_j$, with the red boxes highlighting the shaded time window from (a).
Figure~\ref{fig:resonance}(c) clearly shows that during the shaded time, a doublon can form on site $j=0$. This doublon then persists on that site for about one pumping period, as it can only dissociate at the next crossing of these levels one pumping period later. 
For $U>U_1$, in contrast, the relevant levels do not cross anymore, and this doublon channel is closed. 
As a result, the doublon formation gets strongly suppressed and the pumping behaviour changes entirely [see Fig.~\ref{fig:resonance}(d) and (e)]. In the many-body case, the closing of this channel directly coincides  with the marked increase in $\bar{Q}_R$ shown in the Fig.~\ref{fig:QLR}.

An analogous effect can be found around the second line in Fig.~\ref{fig:QLR} at $U\sim 38J = U_2$, again using a minimal two-particle model, see Fig.~\ref{fig:resonance}(f)-(j). Here, the relevant channels are again two singlons on sites $j$ and $j+2$, but in this case, their energies cross with that of a doublon on the middle site $j+1$. For $U\lesssim U_2$, we again observe the formation of doublons that typically live for one pumping period, and the pumping behavior changes drastically once the doublon channel becomes inaccessible for $U>U_2$.

In both discussed cases, the doublon forms from two particles that are next-to-nearest neighbors and the channels differ by where the doublon forms. There are several other channels starting from nearest neighbors or particles even further apart, but the closing of those channels does not give rise to significant signatures in the many-body case.

\section{Pumping of a single doublon}\label{sec:2particle}
In order to further analyze the differences between the different bands, we now move to a minimal model and explore the pumping of a single doublon.  While the single-particle band structure is fully symmetric in energy, with the top and bottom  $C=1$ bands behaving identically, finite interactions break this symmetry and give rise to different behaviors for a doublon in the top or bottom band.  
We initialize a doublon on a site corresponding to the desired band by tuning the initial phase $\gamma$ and study the resulting pumping dynamics for different repulsive interactions  $U \geq 0$, see Fig.~\ref{fig:2particle}, where the left, middle, and right columns correspond to the lower, middle, and upper bands, respectively.

In the non-interacting case [Fig.~\ref{fig:2particle}~(a,e,i)], the two particles are always pumped independently but identically and hence naturally remain together during the pump, irrespective of the band. The very slight broadening of the left moving cloud stems from the bare tunneling and is also visible in Fig.~\ref{fig:7particles}(a).

In the interacting cases, in contrast, there is a significant difference between the lowest band 
[Fig.~\ref{fig:2particle}~(b,c)] and the other two bands. In the lowest band, the interactions have, for the chosen pump speed, no influence on the pumping and the two particles are always pumped together. In the middle and higher bands, in contrast, the interactions can lead to a dissociation of the doublon and to backscattering, where one particle gets scattered into a lower band and subsequently is pumped into the opposite direction. Interestingly, this asymmetry can lead to a decrease in total energy in this pumped many-body system and is discussed further in Sec.~\ref{sec:enrdk}. 

Similar to the previous discussion, this difference between the bands can be understood by analyzing the energy channels utilized by the two particles during the pumping, shown at the bottom of Fig.~\ref{fig:2particle}. The initial state, i.e., a doublon on site $j=0$, is denoted by the green line. For the lowest band [Fig.~\ref{fig:2particle}(d)], the doublon state first becomes resonant with the state of two neighboring single particles at sites $j=0$ and $1$, plotted as the red dotted line. This state subsequently becomes resonant with the doublon state at site $j=1$ (blue). For that given pumping speed, these two transitions are adiabatic and hence the doublon remains intact.

For a doublon in the middle band, in contrast, the situation is reversed, and the doublon on site $j=0$ directly becomes resonant with a doublon on site $j=-1$, see the Fig.~\ref{fig:2particle}(h). Therefore, the transition is not a sequence of two single-particle transitions but a direct two-particle transition. Since this is not fully adiabatic, a fraction of the doublon initially remains at site $j=0$ and then subsequently dissociates into two neighboring particles (yellow dotted line), of which one now belongs to the lower $C=1$ band. This changes the total energy and, since the lower band is pumped in the opposite direction, gives rise to the observed backscattering. The same mechanism applies to the upper band (right column). Based on our data and analysis, we expect that in a many-body system this process of dissociation of doublons continues until there are no doublons left in higher bands, see below.

We note that for attractive interactions, $U<0$, the situation changes as the doublon states are shifted downwards in energy instead of upwards. This changes the order of the level crossings in the Fig.~\ref{fig:2particle}(d), (h) and (l). As a consequence, now the doublon pump in the upper band becomes stable while isolated doublons in the middle and lower bands dissociate during the pump, transmitting particles to the band above. This occurs because for interaction $U \rightarrow -U$ and phase $\gamma(t) \rightarrow \gamma(t) + \pi$, the energy channels become mirror reflections about zero $E \rightarrow -E$.

\section{Energy decay}\label{sec:enrdk}
\begin{figure}
\centering
    \includegraphics[width=1\linewidth]{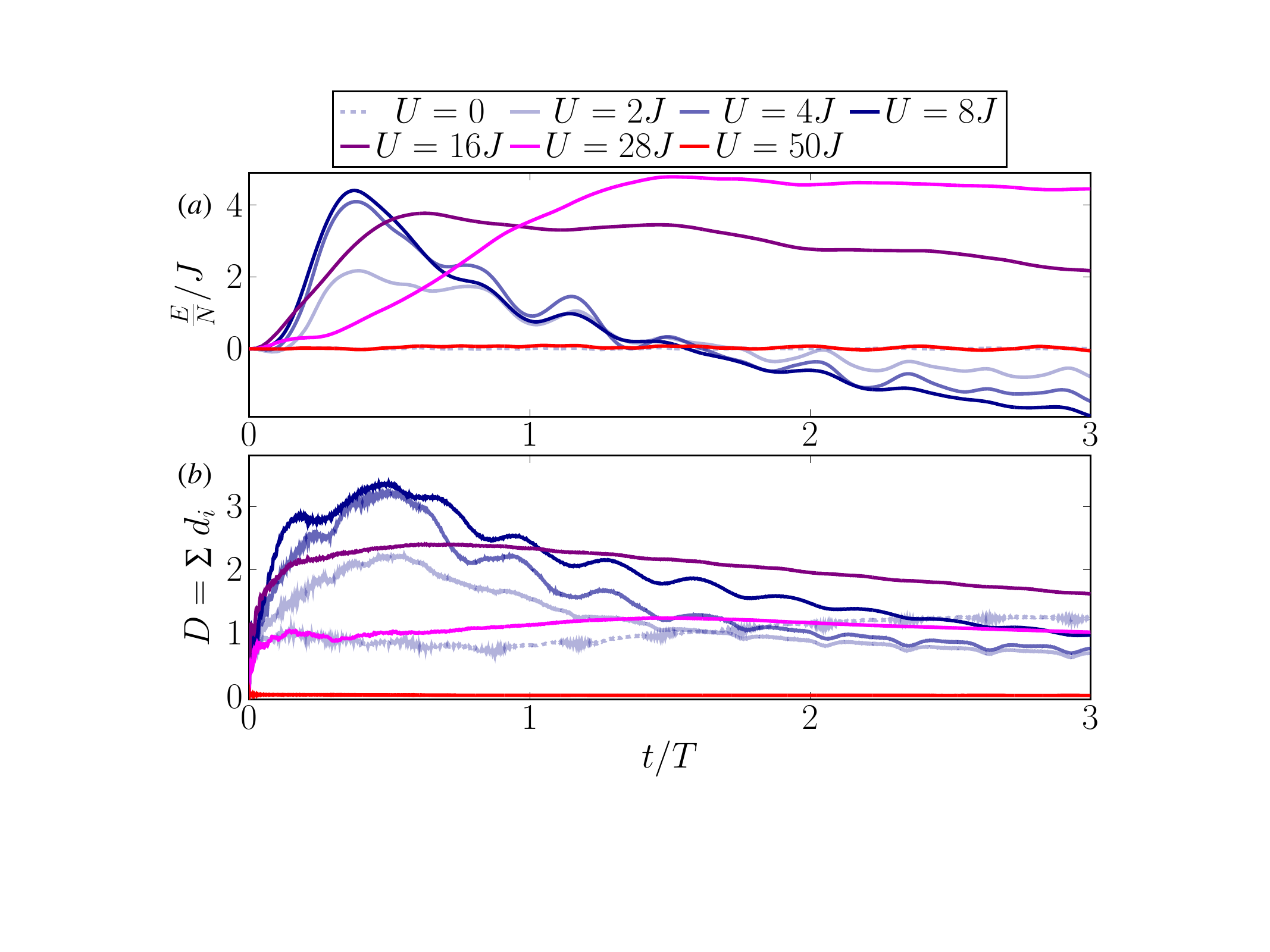}
\caption{Time evolution of (a) total energy and (b) total doublon number  starting from a product state [see Eq.~\eqref{eq:psi0}] with $N=7$ particles and $\Delta = 15J$, averaged over 40 realizations. For the non-interacting ($U=0$) and hardcore boson($U/J=50$) cases, the energy remains essentially constant, while it initially increases in the presence of finite repulsive interactions. Interestingly, for all interactions, the total energy eventually starts to decrease and, for weaker interactions ($U=2J$, $4J$ and $6J$), decreases below the initial energy already during the simulated time.
}
\label{fig:ener}
\end{figure}
In a periodically driven isolated many-body system one normally expects Floquet heating, meaning that its total energy will generically rise with time \cite{Lazarides2014,dAlessio2014} as the system absorbs energy from the drive and heats up.
Here, in contrast, we observe an unusual energy decay. Examples of this energy decay are shown in Fig.~\ref{fig:ener}(a) and highlight that for intermediate interaction strength, $U=2J$, $4J$, and $8J$, the energy drops below the initial energy already on the simulated timescales. These timescales match the times for the decay of the initially created doublons shown in Fig.~\ref{fig:ener}(b), demonstrating that the underlying microscopic mechanism is directly related to the decay of individual doublons discussed in Sec.~\ref{sec:2particle}. 

Starting from the initial product state, bare tunneling generates some doublons during the first tunneling times~\cite{Ronzheimer2013}, as shown in Fig.~\ref{fig:ener}(b), accompanied by an initial increase in the total energy. As discussed in the previous section, for the given pumping speed, doublons in higher bands  are unstable under the pumping, and will dissociate with one particle ending up in a lower band. As a result, the total energy gradually decreases until no doubons remain in higher bands.

\section{Conclusions}
\label{sec:conclusions}
In conclusion, we explored the effect of interactions on the Thouless pump in a quasiperiodic system, the Aubry-Andr{\'e} model with on-site interacting bosons. The time period and the strength of the quasiperiodic potential were chosen such that the two first-order gaps in the single-particle spectrum constitute the relevant energy scale. In the non-interacting limit, this leads to quantized charge pumping in the three first-order subbands with Chern numbers $C=1$, $-2$, and $1$.

We find that the pumped charge starts to deviate from the quantized value already for weak interactions. This result is in sharp contrast with studies of the non-interacting pump in the presence of disorder, where the pump is extraordinarily robust~\cite{Emmanuel2024}. 
For intermediate interactions we identified discrete doublon channels that significantly influence the pumped charge, leading to sharp changes at the points where these channels close. At very large interaction strengths, where doublon formation is strongly suppressed, the system enters into the regime of hard-core bosons and the quantized pump is recovered.

Finally, we demonstrated an interesting asymmetry where the dynamics of an isolated doublon under the pump depends on its initial energy and the sign of the interaction. For repulsive interactions, doublons in the lowest band are pumped stably. In the higher two bands, in contrast, the doublon pump breaks down in the presence of repulsive interactions, and the doublon dissociates by emitting a particle into a lower band. This  gives  rise to a surprising decrease in total energy, contrary to the typical expectation of Floquet heating. For attractive interactions, the role of the bands is reversed and particles get transferred towards higher bands.

Our results demonstrate the richness of pumped quasiperiodic systems and call for more comprehensive studies exploring additional regimes of potential strength and pumping speed in interacting systems. Topological  pumping is not only of fundamental interest, but the ability to deterministically move particles depending on their band is also a central ingredient in recent proposals for quantum computers based on ultracold fermions in optical lattices~\cite{zhu2025, bojovic2025}  and their connectivity could be greatly enhanced by using more sub bands. \\

\textit{Data availability statement:} The data shown in the figures and simulation codes are available at DOI:10.5281/zenodo.18335986~\cite{data}.

\section*{Acknowledgement}
This work was supported by the Deutsche Forschungsgemeinschaft (DFG, German Research Foundation) — 277974659, 436382789, and 493420525 via DFG Research Unit FOR 2414 and large-equipment grants (GOEGrid cluster), the EPSRC Programme
Grant QQQS (EP/Y01510X/1), and the UK Quantum Technology
Hub QCI3 (EP/Z53318X/1). A portion of the numerical simulations were conducted on the MPIPKS HPC cluster.

\bibliography{references}

\end{document}